\documentclass[preprint,showkeys,superscriptaddress,nofootinbib,aps]{revtex4}
 \usepackage{graphicx}
 \usepackage{color} 
\usepackage{amssymb}
 
  \def\tb{\textcolor{black}}
 \begin{document}
 \title{ Charmless $B_{c} \to VV$ decays in the QCD factorization approach}
 \author{Qin Chang}
 \affiliation{Institute of Particle and Nuclear physics,\\
 Henan Normal University, Xinxiang 453007, China}
  \affiliation{Institute of Particle Physics and Key Laboratory of Quark and Lepton Physics (MOE),\\
              Central China Normal University, Wuhan 430079, China}
  \author{Na Wang}
 \affiliation{Institute of Particle and Nuclear physics,\\
 Henan Normal University, Xinxiang 453007, China}
   \affiliation{Institute of Particle Physics and Key Laboratory of Quark and Lepton Physics (MOE),\\
              Central China Normal University, Wuhan 430079, China}
 \author{Junfeng Sun}
 \affiliation{Institute of Particle and Nuclear physics,\\
 Henan Normal University, Xinxiang 453007, China}
 \author{Lin Han}
 \affiliation{Institute of Particle and Nuclear physics,\\
 Henan Normal University, Xinxiang 453007, China}
  \begin{abstract}
In this paper, we studied the charmless $B_{c}\to VV$ ($V$ denotes  the light ground $\rm SU(3)$ vector meson) decays within the framework of QCD factorization. In the evaluation, two different schemes for regulating the end-point divergence are adopted. One (scheme I) is to use parameterization model, which is usually employed in the QCD factorization approach; the other (scheme II) is based on the infrared finite gluon propagator of Cornwall prescription. It is found that, in the annihilation amplitudes, the end-point divergence appears only in the power-suppressed corrections related to the twist-3 distribution amplitudes of $V$-meson. The strength of annihilation amplitudes evaluated in scheme II is generally larger than the one in scheme I. Numerically, in the decay modes considered in this paper,  the CKM-favored $B_{c} \to \rho^{-}\omega, K^{*-}K^{*0} $ decays have the relatively large branching fractions, $\sim {\cal{O}}(10^{-7})$, and hence are hopeful to be first observed by the future experiments. In addition, all of the decay modes are dominated by the longitudinal polarization state; numerically,  $f_{L}(B_{c} \to VV) \gtrsim 99\%$.
  \end{abstract}
 \maketitle

 \section{Introduction}
 \label{sec1}
The  $B_{c}^-$  meson is the only ground-pseudoscalar consisting of two heavy quarks with different flavor, namely a $\bar{c}$ and a $b$ quark.
The difference of components flavors forbids $B_{c}^-$  meson to annihilate into gluons or photons through strong interactions or electromagnetic interactions. Moreover, the $B_{c}$  meson lies below the $BD$ threshold. Therefore,  it is considerably more stable than the charmonium or bottomonium states, and decays mainly through weak interaction. Since the $b$ and $c$ quarks can decay individually, the $B_{c}$ meson has much richer  decay modes than  $B_{u,d,s}$ mesons~\cite{Chang:1992pt}, that could provide an ideal ground for studying the hadronic weak decays of heavy flavor quarks.

 In the standard model~(SM), the $B_{c}$ weak decays can be divided into three categories: (1) the $b \to (c,u)W^{-}$ process with $\bar{c}$-quark as a spectator; (2) the $\bar{c} \to (\bar{s},\bar{d})W^{-}$ process with $b$-quark as a spectator; (3) the pure weak annihilation $b\bar{c} \to W^{-}$ transition. Among the multitudinous $B_{c}$ decay modes, the pure weak annihilation decay channels are expected to take $10\%$ shares~\cite{Brambilla:2004wf}. In the pure annihilation $B_{c}$ decays, the major part comes from the ``tree" annihilation processes induced by the  CKM-favored $B_{c}^{-}\to s\bar{c}$ transition because of the sizable $c$-quark mass, while the charmless annihilation decays are relatively rare due to the power-supression.

Experimentally, the production of $B_{c}$ meson in hadron collisions implies the simultaneous production of $b\bar{b}$ and $c\bar{c}$ pairs, and  therefore is relatively rarer than the other $b$ mesons~\cite{Brambilla:2010cs}.
The heavy $B_{c}$  meson was first observed by CDF collaboration from Run-I at Tevatron through the semileptonic decay mode $B_{c}^- \to J/\Psi l^- \bar{v}$~\cite{Abe:1998wi}. At the Large Hadron Collider (LHC) with a luminosity of about ${\cal L}=10^{34}{\rm cm}^{-2}{\rm s}^{-1}$,  around $5\times 10^{10}$ $B_c$ events can be produced  per year~\cite{Gouz:2002kk}, and the measurements of the mass and lifetime of  $B_{c}$ meson have reached a very precise degree, for instance, $m_{B_{c}}=6276.28\pm1.44\pm0.36$ MeV~\cite{Aaij:2013gia} and $\tau_{B_{c}}=513.4\pm11.0\pm5.7\,{\rm fs}$ ~\cite{Aaij:2014gka} reported by the LHCb collaboration. Benefiting from the large production rate at LHC, a lot of $B_{c}$ meson decays have been observed by LHCb collaboration, for instance: the $B_{c}^{+}\to J/\Psi \pi^{+}\pi^{-}\pi^{+}$~\cite{LHCb:2012ag}, $\Psi(2S)\pi^{+}$~\cite{Aaij:2013oya}, $ J/\Psi D_{s}^{(*)}$~\cite{Aaij:2013gia}, $J/\Psi K^{+}$~\cite{Aaij:2013vcx}, $ J/\Psi K^{+} K^{-} \pi^{+}$~\cite{Aaij:2013gxa} and $D^{0} K^{+}$~\cite{Aaij:2017kea} decay modes induced by the $b$ quark decay, the first $c$ quark decay mode $B_{c}^{+} \to B_{s}^{0}\pi^{+}$~\cite{Aaij:2013cda} and the baryonic decay mode $B_{c}\to J/\Psi p\bar{p}\pi^{+}$~\cite{Aaij:2016xxs} {\it etc.}. In the near future, more $B_{c}$ weak decays are expected to be measured at LHC with its high collision energy and high luminosity.

Theoretically, the weak decays of $B_c$ meson are generally complicated because of its heavy-heavy nature and the participation of strong interaction, but they also provide opportunities to study the perturbative and nonperturbative QCD, final state interactions and heavy quarkonium properties, {\it etc.}.
In the past years, some theoretical investigations have been carried out on the properties of $B_c$ meson decays based on the QCD-inspired approaches, for instance, the operator product expansion~\cite{Bigi:1995fs,Beneke:1996xe}, the QCD sum rule~\cite{Kiselev:2003ds,Kiselev:2000pp}, the nonrelativistic QCD~\cite{Bodwin:1994jh}, the pQCD factorization approach~\cite{Du:1991np,Yang:2010ba,Sun:2014ika,Sun:2014jka,Liu:2009qa,Rui:2015iia,Rui:2016opu,Liu:2010nz,Liu:2010da,Liu:2010kq,Xiao:2011zz,Liu:2017cwl}, QCD factorization~(QCDF)~\cite{Sun:2015exa,DescotesGenon:2009ja,Wang:2016qli}, the QCD relativistic potential models~\cite{Colangelo:1999zn,Kiselev:2000jc} and the Bethe-Salpeter method~\cite{Ju:2014oha,Chang:2014jca}.
The two-body non-leptonic charmless $B_c$ decay can occur only via the weak annihilation diagrams: the $b$ and $c$ quarks annihilate into a charged $W^\pm$ boson that decays into a pair of a $u$ and a $d/s$ quark, which further hadronize into the two light mesons. Therefore, the charmless  $B_{c} \to M_1M_2$ ($M_{1,2}$ are the light mesons) decays are very suitable for probing the strength of annihilation contribution and
and exploiting the related mechanism,  which are currently important issues in the $B$ physics. Recently, the charmless $B_{c} \to VV$ decays are studied by using the $SU(3)$ flavor symmetry~\cite{DescotesGenon:2009ja} and the pQCD approach~\cite{Liu:2009qa}. In this article, we will revisit these decay modes by employing the QCDF approach~\cite{Beneke1,Beneke2} to cope with the hadronic matrix elements.

In the theoretical framework based on the collinear factorization, the calculation of weak annihilation amplitude always suffers from the end-point singularities. In practice, there are two different phenomenological schemes proposed to deal with the end-point divergence in the QCDF approach.
The scheme I is the parameterization method and has been widely employed in the previous works. In this scheme,  the divergent integral is regulated by performing cutoff at $x=\Lambda_h/m_b$, where $x$ is the momentum fraction of quark and $\Lambda_h\sim \Lambda_{\rm QCD}$ is
the parameter characterizing the point of cutoff (typically, $\Lambda_h=0.5{\rm }\mathrm{GeV}$~\cite{Beneke:2001ev,Beneke:2003zv})  ; meanwhile, the integrals near end-point are treated as signs of infrared sensitive contributions, and parameterized by introducing the phenomenological parameters $\rho_{A}$ and $\phi_{A}$. Explicitly, the divergent integral is parameterized as $\int^{1}_{0}dx/x \to X_{A}={\rm ln}(m_b/\Lambda_h)\,(1+\rho_{A} e^{i\phi_{A}})$~\cite{Beneke:2001ev,Beneke:2003zv}.
As an alternative to the parameterization method, the end-point divergence could also be regulated by introducing an infrared finite dynamical gluon propagator~\cite{Cornwall:1981zr,Cornwall:1989gv,Papavassiliou:1991hx}, namely scheme II, which also has been successfully applied to the  nonleptonic $B_{u,d,s}$ meson decays~\cite{BarShalom:2002sv,SuFang,Chang:2008tf,Chang:2012xv,ChangQin}.  In this paper, above two regulation schemes are adopted respectively in our evaluation.

Our paper is organized as follows. In section~\ref{sec2}, after a brief review of the theoretical framework for the two-body charmless hadronic $B_{c}$ decays, the detailed calculation and discussion for the annihilation amplitudes in the QCDF are presented. Section~\ref{sec3} is devoted to the numerical results and discussion. Finally,  we summarize in Sec.~\ref{sec4}. The explicit expressions for the decay amplitudes and the relevant input parameters are collected in appendixes~\ref{appendix A} and \ref{appendix B}, respectively.

 \section{Theoretical framework and calculation}
 \label{sec2}

 \subsection{The effective weak Hamiltonian and hadronic matrix element}
 \label{sec2.1}

The effective weak Hamiltonian responsible for the charmless $B_{c}^{-}\to V_{1}V_{2}$ decays can be written as~\cite{Buchalla:1995vs,Buras:1998raa}
 \begin{equation}
 {\cal H}_{\rm eff}=\frac{G_{F}}{\sqrt{2}}\,V_{cb}V_{up}^{\ast}\,\Big[C_{1}(\mu)Q_{1} + C_{2}(\mu)Q_{2}\Big] + {\rm h.c.},
 \label{eq:Heff}
 \end{equation}
where $G_{F}$ is the Fermi coupling constant, $V_{cb}V_{up}^{*}$~($p=d,s$) is the product of CKM matrix elements~\cite{CKM}, and $Q_{1,2}$ are local four-quark operators arisen from $W$-boson exchange and defined as
\begin{eqnarray}
 Q_{1} &= \left[\bar{c}_{\alpha}\gamma^{\mu}(1-\gamma_{5})b_{\alpha}\right] \left[\bar{p}_{\beta}\gamma_{\mu}(1-\gamma_{5})u_{\beta}\right]\,, \nonumber \\
 Q_{2} &= \left[\bar{c}_{\alpha}\gamma^{\mu}(1-\gamma_{5})b_{\beta}\right] \left[\bar{p}_{\beta}\gamma_{\mu}(1-\gamma_{5})u_{\alpha}\right]\,,
 \label{eq:Qi}
\end{eqnarray}
with the  color indices of  $\alpha$ and $\beta$. The Wilson coefficient $C_{i}(\mu)$ in  Eq.~(\ref{eq:Heff}) describes the coupling strength for a given operator and summarizes the physical contributions above scale of $\mu$.  They are calculable perturbatively with the
renormalization group  improved perturbation theory \cite{Buchalla:1995vs,Buras:1998raa}.  In addition,  the $\overline{\rm MS}$ renormalization scheme~(RS) is employed in this work.

\begin{figure}[t]
\includegraphics[width=10.0cm]{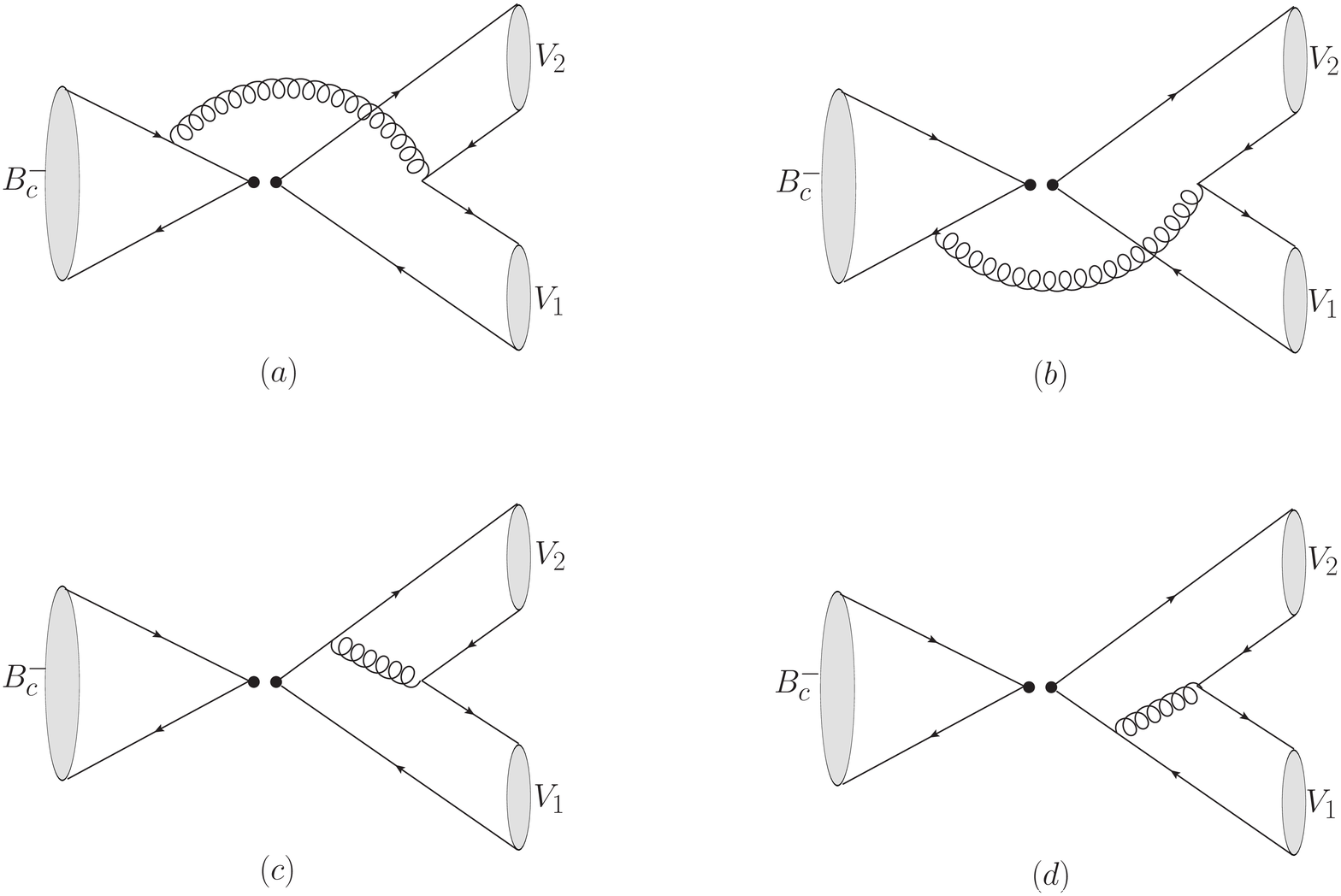}
\caption{The Feynman diagrams for the charmless $B_{c}^{-}\to V_{1}V_{2}$ decays at the order of $\alpha_s$.}
\label{fig.1}
\end{figure}

In order to obtain the decay amplitudes, the remaining works are to accurately calculate the hadronic matrix elements of local operators, $\langle VV | Q_{i}(\mu) | B_{c} \rangle$. In the QCDF, following the prescription proposed in Ref.~\cite{Lepage:1980fj}, the hadronic matrix elements for the pure annihilation $B\to M_1M_2$ decay can be written as the convolution integrals of the scattering kernel with the distribution amplitudes~(DAs) of the participating mesons~\cite{Beneke1},
  \begin{eqnarray}\label{eq:fe}
  \langle M_1M_2|Q_i|\bar{B}\rangle &=& f_{B}f_{M_{1}}f_{M_{2}}\int d x \,d y \,d z\,{\cal T}_{i}^{II}(x,y,z)\,\varphi_{M_1}(x)\,\varphi_{M_2}(y)\,\varphi_{B}(z)\,,
\label{element}
\end{eqnarray}
where $x\,,y\,,z$  are the momentum fractions; $f_{B}$ and $f_{M}$ are decay constants of the $B$ and light mesons,  respectively; and the kernel ${\cal T}_{i}^{II}(x,y,z)$ is hard-scattering functions.

For the  $B_{c}\to VV$ decays, the kernel ${\cal T}_{i}^{II}(x,y,z)$ at the order of $\alpha_s$ can be obtained by calculating the  Feynman diagrams shown in  Fig.~\ref{fig.1}, in which Figs.~(a), (b) and Figs.~(c), (d) are non-factorizable  and factorizable topologies, respectively. In these topologies, the contributions of factorizable diagrams, Figs.~(c) and (d), cancel each other exactly in the QCDF approach due to the conservation of the vector current, partial conservation of axial-vector current, and the approximation that the twist-2 and twist-3 distribution amplitudes for the final states, $V_1$ and $V_2$, have the same asymptotic expression. This situation is the same as the case of $B_{u,d,s}\to MM$ decays~\cite{Beneke:2001ev,Beneke:2006hg,Beneke:2003zv} and $B_{c}\to PP\,,PV$ decays~\cite{Wang:2016qli}. In addition, because of the mismatch of the color indices, there is no contribution with insertion of the color-singlet operator, $Q_1$,  at the order of $\alpha_s$.


Applying the QCDF formula, the decay amplitudes of $B_{c}\to VV$ decays can then be written as
\begin{equation}
 \langle V_{1}V_{2}|{\cal H}_{\rm eff}|B_{c}^{-}\rangle^{\lambda} \propto f_{B_{c}}f_{V_{1}}f_{V_{2}}\,b_{2}^{\lambda}(V_{1},V_{2})\,,
 \label{eq:matrixelement}
\end{equation}
where $\lambda=0, \pm$ denote the helicities of the final-state vector mesons. The effective coefficient $b_{2}^{\lambda}(V_{1},V_{2})$ is defined as~\cite{Beneke:2006hg,Beneke:2001ev}
\begin{equation}
b_{2}^{\lambda}(V_{1},V_{2})=\frac{C_{F}}{N_{c}^{2}}\,C_{2}\,A_{1}^{i,\lambda}(V_{1},V_{2})\,,
\end{equation}
where $C_{F}=4/3$ with $N_{c}=3$. The superscript `$i$' on $A_{1}^{i,\lambda}$ refers to the gluon emission from the initial-state quarks, and the subscript `$1$' refers to the $(V-A)\otimes(V-A)$ Dirac structure of the inserted four-quark operator $Q_2$.
The the explicit expressions of the building blocks, $A_{1}^{i,\lambda}$, will be given in the following subsections.

\subsection{$A_{1}^{i,\lambda}(V_{1},V_{2})$ in scheme~I}
\label{sec2.2}
As aforementioned,  the annihilation amplitude always suffers from the end-point divergence in the QCDF approach. Traditionally, the divergence is usually parameterized by introducing the complex parameters, $X_{A}={\rm ln}(m_b/\Lambda_h)\,(1+\rho_{A} e^{i\phi_{A}})$~\cite{Beneke:2001ev}, in which the phenomenological parameters $\rho_{A}$ and $\phi_{A}$ reflect the strength and strong phase of the annihilation contributions. These parameters can only be obtained by fitting to the well-measured $B$ decay modes, and then extended to predict the other decays~\cite{Wang:2013fya,Chang:2015wba,Chang:2014yma,Chang:2014rla}. Despite the fact that such a treatment is not entirely self-consistent, it is nevertheless useful for estimating the annihilation amplitude for particular final states, and has been wildly used in the theoretical calculation.

 In this subsection, following such parameterization method, we adopt a similar way to estimate the charmless annihilation $B_{c} \to VV$ decays. Given that $m_{B_{c}}\simeq m_{b}+m_{c}$, the $B_c$ meson can be approximated as a non-relativistic~(NR) bound state that is dominated entirely by the two-particle Fock state built by a $\bar{b}$ and a $c$ quark. In such a NR limit, the soft components of the heavy-quark momentum can be neglected, and we can set the momentum of the valence quark to $p_{b}^{\mu}=m_{b}v^{\mu}$ and $p_{c}^{\mu}=m_{c}v^{\mu}$, where $v^{\mu}$ is the four-velocity of the $B_c$ meson. This means that the light-cone momenta of the quarks are fixed according to their masses, and the distribution amplitude of the $B_c$  meson then takes  the peak form $\Phi_{B_c}(z)\propto \delta(z-m_c/m_{B_c})$~\cite{Bell:2008er,Du:1991np,Brodsky:1985cr}.

 Following the convention adopted in Ref.~\cite{Beneke:2006hg} and using the peak form of $\Phi_{B_c}(z)$, we obtain  the longitudinal component of  annihilation amplitudes written as
 \begin{eqnarray}
A_{1}^{i,0}(V_{1},V_{2}) &=& \pi\alpha_{s}(\mu)\int_0^1\! dx dy \, \Bigg\{\Phi_{V_{1}}(x)\,\Phi_{V_{2}}(y)\,\bigg[\frac{1}{x\big[(x+\bar{y})z_{b}-{x}\bar{y}-i\epsilon\big]}
- \frac{1}{\bar{y}\big[({x}+\bar{y})z_{c}-{x}\bar{y}-i\epsilon\big]}\bigg] \nonumber \\
&& - r_{\chi}^{V_{1}}\,r_{\chi}^{V_{2}}\,\Phi_{v_{1}}(x)\,\Phi_{v_{2}}(y)\, \bigg[\frac{{x}\bar{y}+({x}+\bar{y}-2{x}\bar{y})z_{b}}{{x}\bar{y}\big[({x}+\bar{y})z_{b}-{x}\bar{y}-i\epsilon\big]}
 - \frac{{x}\bar{y}+({x}+\bar{y}-2{x}\bar{y})z_{c}}{{x}\bar{y}\big[({x}+\bar{y})z_{c}-{x}\bar{y}-i\epsilon\big]}\bigg]\Bigg\}\,;\nonumber \\
 \label{eq:a1i0_schemeI}
\end{eqnarray}
and the transverse components  are
\begin{eqnarray}
A_{1}^{i,-}(V_{1},V_{2})& = &\pi\alpha_{s}\tb{(\mu)}\frac{2m_{1}m_{2}}{m_{B_{c}}^2}\int_0^1\! dx dy \,
\Bigg\{\phi_{b1}(x)\,\phi_{b2}(y)\,\bigg[\frac{\bar{y}+z_{b}}{x\bar{y}\big[(x+\bar{y})z_{b}-{x}\bar{y}-i\epsilon\big]}
 \nonumber \\[0.2cm]
& & +\frac{\bar{x}}{x^2\big[({x}+\bar{y})z_{b}-{x}\bar{y}-i\epsilon\big]}
 - \frac{\bar{x}\bar{y}}{{x}\big[(x+\bar{y})z_{b}-x\bar{y}-i\epsilon\big]^2}
\nonumber \\[0.2cm]
& &-\frac{z_{c}}{x\bar{y}\big[({x}+\bar{y}){z_{c}}-{x}\bar{y}-i\epsilon\big]}
 +\frac{\bar{x}}{\big[(x+\bar{y}){z_{c}}-x\bar{y}-i\epsilon\big]^2}\bigg]\Bigg\}\,,
 \label{eq:a1im_schemeI}
\end{eqnarray}
\begin{eqnarray}
A_{1}^{i,+}(V_{1},V_{2})& = &\pi\alpha_{s}\tb{(\mu)}\frac{2m_{1}m_{2}}{m_{B_{c}}^2}\int_0^1\! dx dy  \,
\Bigg\{\phi_{a1}(x)\,\phi_{a2}(y)\,\bigg[\frac{z_{b}}{x\bar{y}\big[(x+\bar{y})z_{b}-{x}\bar{y}-i\epsilon\big]}
 \nonumber \\[0.2cm]
& &-\frac{y}{\big[(x+\bar{y})z_{b}-x\bar{y}-i\epsilon\big]^2}
 - \frac{x+z_{c}}{x\bar{y}\big[({x}+\bar{y}){z_{c}}-{x}\bar{y}-i\epsilon\big]}
\nonumber \\[0.2cm]
& &- \frac{y}{\bar{y}^2\big[({x}+\bar{y}){z_{c}}-{x}\bar{y}-i\epsilon\big]}
 +\frac{xy}{\bar{y}\big[(x+\bar{y}){z_{c}}-x\bar{y}-i\epsilon\big]^2}\bigg]\Bigg\}\,,
 \label{eq:a1ip_schemeI}
\end{eqnarray}
\tb{where $x$~($\bar{x}\equiv1-x)$ and $y$~($\bar{y}\equiv1-y$) are the longitudinal momentum fractions of (anti-) quarks in $V_{1}$ and $V_{2}$ mesons, respectively}; $z_{b}$ and $z_{c}$ denote the relative sizes of the $b$- and $c$-quark masses, $z_{b}=m_{b}/m_{B_c}$ and  $ z_{c}=m_{c}/m_{B_c}$; \tb{$m_1$ and $m_2$ are the masses of the final-state vector mesons;} and the factor $r_{\chi}^{V}$ is defined as
\begin{equation}
 r_\chi^V(\mu) = \frac{2 m_{V}}{m_{B_c}}\,\frac{f_{V}^{\perp}(\mu)}{f_V}\,
\end{equation}
\tb{where $m_{V}=m_1\,,m_2$; $f_{V}^{\perp}(\mu)$ is the scale-dependent transverse decay constant.} Finally, we checked with the full results that in the limit  $z_b \to 1$ and $z_{c}\to 0$ coincide with the results for $B_{u,d,s}\to VV$ decay in the heavy quark limit given by Eqs. (A.17) and (A.18) in Ref.~\cite{Beneke:2006hg}. \tb{In our following evaluations, the $c$-quark mass is reserved; in fact, we will show in the follows that the unnegligible  $c$-quark masse plays an important role for eliminating the end-point divergency in the amplitudes of twist-2 part. }

For the longitudinal amplitude, Eq.~(\ref{eq:a1i0_schemeI}), only a few signs change in comparison with the known results for $B_{c}\to PP$ or $PV$ decays~\cite{Wang:2016qli}. Because  $r_\chi^V(\mu) $ is suppressed by one power of $\mathrm{\Lambda_{QCD}}/m_b$,
the contributions related to the twist-3 DAs in Eq.~(\ref{eq:a1i0_schemeI}) are \tb{small} numerically.
For the transverse amplitudes,  from Eqs.~(\ref{eq:a1im_schemeI}) and (\ref{eq:a1ip_schemeI}), one can find that only the twist-3 terms of the light-cone projection operator contribute to them, and the transverse amplitudes are suppressed by two powers of $\mathrm{\Lambda_{QCD}}/m_b$ compared with the longitudinal amplitude. Therefore, the $B_{c}\to VV$ decay is expected to be dominated by longitudinal polarization.

Using the asymptotic expression for the distribution amplitudes of light vector meson~\cite{Beneke:2006hg,Beneke:2001ev,Ball:1998sk}
\begin{eqnarray}\label{asw1}
 &\Phi_{V}(x)=\phi_{\perp}^V(x)=6x(1-x)\,,\quad \phi_{a}(x)=\phi_{b}(\bar{x})=3\bar{x}^2\,,\quad \Phi_{v}(x)=3(x-\bar{x})\,,
\label{asw2}
\end{eqnarray}
the weak annihilation amplitudes of $B_{u,d,s}\to VV$ decays exhibit logarithmic and even linear infrared divergences~\cite{Beneke:2006hg}, hence the analyses of these decays suffer from large uncertainties. It \tb{should be noted} that the integral of the twist-2 part encounters the end-point divergence for  $B_{u,d,s}\to VV$ decay, \tb{but }is finite for  $B_{c}\to VV$ decay due to the sizable $c-$quark mass which results in a complex contribution~\tb{(it can be clearly seen from the second term proportional to $\Phi_{V_{1}}\Phi_{V_{2}}$ in Eq.~(\ref{eq:a1i0_schemeI})).}
Unfortunately,   the logarithmic divergence \tb{ exists still} at twist-3 level for $B_{c}\to VV$ decay. Further considering the fact that all of the twist-3 contributions are   power-suppressed by $(\mathrm{\Lambda_{QCD}}/m_b)^2$ relative to the twist-2 part, we can expect that  the prediction for $B_{c}\to VV$ decay in the framework of  QCDF  should be much more precise than $B_{u,d,s}\to VV$ decays.

In the numerical evaluation, one will encounter the physical-region singularity of the on mass-shell quark propagators and endpoint divergence of the gluon propagators in Eqs.(\ref{eq:a1i0_schemeI})-(\ref{eq:a1ip_schemeI}). Here, we adopt the Cutkosky rule to deal with the singularities~\cite{Su:2006uy,Cutkosky:1960sp}.
For the divergence arising from the gluon propagator in the twist-3 part, because those terms are complex and hardly to be expressed as polynomial of $X_{A}$, \tb{
we take the integral interval of \tb{${x},y \in [ \Lambda_h/m_b, 1 ]$}. In the following numerical evaluations of scheme I, we use $\Lambda_h=\Lambda_{{\rm QCD}}^{\overline{\rm MS}\,,n_f=3}=332~{\rm MeV}$~\cite{PDG2016}, which is a little smaller than the typical choice, $\Lambda_h=500~{\rm MeV}$. The effect of $\Lambda_h$ will be discussed briefly in the follows. }

\tb{The numerical results for the building blocks $A_{1}^{i,\lambda}$ with the default inputs summarized in Appendix B
are}
\tb{
\begin{eqnarray}
 A_{1}^{i,0}(V_{1},V_{2}) & = & \pi \left[(-5.64 - 6.22i) - r_{\chi}^{V_{1}}r_{\chi}^{V_{2}}(-1.83 - 2.94i)\right],  \\[0.2cm]
 A_{1}^{i,-}(V_{1},V_{2}) & = &\pi \frac{2\,m_{1}\,m_{2}}{m_{B_{c}}^2}(3.32 + 4.57i),\\[0.2cm]
 A_{1}^{i,+}(V_{1},V_{2}) & = &\pi \frac{2\,m_{1}\,m_{2}}{m_{B_{c}}^2}(-10.50 - 1.49i)\,.
 \label{eq:a1i_schemeI_num1}
\end{eqnarray}
}
As we expected, the transverse amplitudes and the twist-3 term in the longitudinal amplitude are numerically \tb{small} due to the power-suppression factor \tb{${m_{1}m_{2}}/{m_{B_{c}}^2}\sim {\cal O}(10^{-2})$}, and therefore, the amplitude of $B_{c}\to VV$ decay is dominated by the contribution of twist-2 term in $A_{1}^{i,0}$. \tb{Further considering the fact that the value of $\Lambda_h$ only affects the integral of twist-3 part, we can conclude that the effect of $\Lambda_h$ on the total amplitude is small. For instance,  using $\Lambda_h=0.2$ and $0.5~{\rm GeV}$, respectively, we obtain
\begin{eqnarray}
A_{1}^{i,0}(\rho^-\,\omega)= -17.58-19.28i
 \quad {\rm vs.}
\quad -17.51-19.28i
 \end{eqnarray}
 for  $B_c\to\rho^-\,\omega$ decay. It can be clearly seen that the theoretical uncertainty induced by $\Lambda_h$ is at the level of $\lesssim1\%$.}

\subsection{$A_{1}^{i,\lambda}(V_{1},V_{2})$ in scheme~II}
\label{sec2.3}
 In this subsection, we shall quote the infrared finite gluon propagator to regulate the divergences in the annihilation amplitudes. The infrared finite~(IR) dynamical gluon propagator, which is shown to be not divergent as fast as ${1}/{q^2}$, has been successfully applied to various hadronic $B_{u,d,s}$ decays~\cite{BarShalom:2002sv,SuFang,Chang:2008tf,Chang:2012xv,ChangQin}. It should be noted that \tb{an IR finite gluon propagator typically leads to a freezing coupling $\alpha_s(0)$~\cite{Deur:2016tte,Aguilar:2002tc}~(one may refer to Ref.~\cite{Deur:2016tte} for detail)}. The infrared finite behavior is not only obtained from solving the well-known Schwinger-Dyson equation~\tb{\cite{Cornwall:1981zr,Binosi:2016nme}}, but also supported by recent lattice simulations~\cite{Aouane:2012bk,Gongyo:2013sha} and the studies based on the light-front holographic (AdS$_5$) QCD~\cite{Deur:2016cxb}. \tb{In addition,  a freezing $\alpha_s$ is also used as a regulator in Ref.~\cite{Courtoy:2013qca} as we do in this scheme.}

In the practice, we adopt the Cornwall's prescription for the IR finite  gluon propagator~\cite{Cornwall:1981zr},
\begin{equation}
 D(q^{2})=\frac{1}{q^2-M_{g}^2(q^2)+i\epsilon}\,,
 \label{eq:gluonprop}
\end{equation}
where $q^2$ denotes the gluon momentum squared. The corresponding coupling constant \tb{ including the quark loops correction }reads~\tb{\cite{Cornwall:1981zr,Cornwall:1989gv,Papavassiliou:1991hx}}
\tb{
\begin{equation}\label{eq:cornwall1}
 \alpha_{s}^{\rm C}(q^2)=\frac{12\pi}{33\ln\left[\frac{q^2+\epsilon M_{g}^2(q^2)}{\Lambda_{\rm C}^2}\right]-2n_{f}\ln\left[\frac{q^2+\epsilon M^2}{\Lambda_{\rm C}^2}\right]}\,,
\end{equation}
where $\Lambda_{\rm C}$ is the QCD scale, $\epsilon= 4.8$, $M= 0.42\,{\rm GeV}$  is identified with the string tension ~\cite{Papavassiliou:1991hx,Deur:2016tte}, and $n_{f}$ is the number of active quark flavors at a given scale.}  The dynamical gluon mass $M_{g}^{2}(q^{2})$ is given by~~\tb{\cite{Cornwall:1981zr,Cornwall:1989gv,Papavassiliou:1991hx}}
\tb{
\begin{equation}
 M_{g}^{2}(q^{2}) = m_{g}^2 \left[\frac{\ln(\frac{q^{2}+4m_{g}^{2}}{\Lambda_{\rm C}^{2}})}
{\ln(\frac{4m_{g}^{2}}{\Lambda_{\rm C}^{2}})}\right]^{-\frac{12}{11}}\,,
\label{eq:effmass}
\end{equation}
where $m_{g}$ is the effective gluon mass scale with a typical value $m_{g}=0.5\pm0.2~{\rm GeV}$~\cite{Cornwall:1981zr}. The value of $m_{g}$ can be determined from the phenomenological information. For instance, a good description of the experimental pion and kaon form factors is obtained for $m_{g}=0.54~{\rm GeV}$~\cite{Ji:1986uh,Ji:1989cb};  while,  the authors of Ref.~\cite{Aguilar:2001zy} find that $m_{g}=0.70~{\rm GeV}$ describes the pion form factor data well; the value $m_{g}=0.44~{\rm GeV}$ is suggested to analyze the photon-to-pion transition form factor and   $\gamma\gamma\to\pi^+\pi^-$ decay~\cite{Brodsky:1997dh};  the similar values are obtained by fitting to the experimental data of non-leptonic  $B$ decays,  $m_{g}=0.5\pm0.05~{\rm GeV}$ for $B_{u,d}$ decays~\cite{Chang:2008tf} and $m_{g}=0.48\pm0.02~{\rm GeV}$ for $B_s$ decays~\cite{Chang:2012xv}. One may refer to Ref.~\cite{Deur:2016tte} and literatures therein for details for this part. In this work, we take a conservative choice $m_{g}=0.5\pm0.2~{\rm GeV}$~\cite{Cornwall:1981zr}.
}

\tb{
The typical value of $\Lambda_{\rm C}$ is  $0.26\pm 0.05 {\rm GeV}$~\cite{Cornwall:1981zr,Cornwall:1989gv,Papavassiliou:1991hx,Deur:2016tte}, which is usually used for, for instance,  studying the IR behavior of strong coupling.
For the annihilation $B_c$ decays, the range of momentum squared of gluon is very large;
therefore, the values of $\Lambda_{\rm C}$, as well as $n_f$, should be non-universal in different $q^2$ bins in principle. In addition,  the large momentum-transfer dependence of the coupling $\alpha_s$ is generally specified by perturbative QCD (PQCD) and its renormalization group equation. Thus, in order to obtain the values of  $\Lambda_{\rm C}$, we try to match $\alpha_s^{\rm C}$ to $\alpha_s^{\rm PQCD}$ at $q^2=m_b^2\,,m_c^2$ and $1\,{\rm GeV^2}$ with $n_f=5\,,4$ and $3$, respectively.
}

\tb{
In the matching procedure, the world averages of $\Lambda_{\rm PQCD}^{(n_f)}$~\cite{PDG2016} listed in Table~\ref{tab:fit} are used;  in addition, the approximate analytical expression for $\alpha_s^{\rm PQCD}$ up to order $\beta_3$~\cite{Chetyrkin:1997sg} in the $\overline{\rm MS}$ RS is employed.  Using the matching condition  $\alpha_s^{\rm C}=\alpha_s^{\rm PQCD}$, we finally obtain
\footnote{ \tb{These values can be treated as the ``effective" scale absorbing the higher order loops ``corrections" because the 4-loops result for $\alpha_s^{\rm PQCD}$ is used in the matching procedure~(In other words, we require $\alpha_s^{\rm C}$ with a proper $\Lambda_{\rm C}$ to reproduce the 4-loops $\alpha_s^{\rm PQCD}$ in $\overline{\rm MS}$ RS at interval of $q^2>1 {\rm GeV^2}$ ). } }
\begin{equation}
\Lambda_{\rm C}^{n_f=3\,,4\,,5}[{\rm GeV}]=0.35\,,0.22\,,0.10\, ,
\end{equation}
 which are in agreement with the typical value $0.26\pm 0.05\, {\rm GeV}$ except at large $q^2>m_b^2$ with $n_f=5$. Such values will be used in the intervals $q^2< m_c^2$, $m_c^2<q^2< m_b^2$ and $q^2>m_b^2$
\footnote{\tb{In the QCDF, the pole mass of the light quarks, $u\,,d$ and $s$, are taken to be zero in the heavy quark limit, therefore the case of $n_f=2$ is not considered. The effect of such approximation is trivial numerically because it corresponds to a very narrow integral space.}  }
, respectively, in our following evaluations.   }

\begin{table}[t]
\caption{\label{tab:fit} \tb{ The results of $\alpha_s$ with the scales $\Lambda$ at different $q^2$ . The uncertainties in the last  row for $\alpha_s^{\rm C}(q^2)$ is induced by $m_{g}=0.5\pm0.2~{\rm GeV}$.  See the text for the further explanation.  }}
\renewcommand*{\arraystretch}{1.1}
\begin{center}\setlength{\tabcolsep}{5pt}
\begin{tabular}{lccccccccc}
\hline\hline
 $q^2$[{\rm GeV}]& $0$&$1$&$m_c^2$ & $m_b^2$ &$m_Z^2$\\ \hline
$n_f$&$3$&$3$ &$4$ &$5$ &$5$\\\hline
$\Lambda_{\rm PQCD}[{\rm GeV}]$~\cite{PDG2016}&---&$0.332$ &$0.292$ &$0.210$ &$0.210$\\
$\Lambda_{\rm C}[{\rm GeV}]$&$0.35$&$0.35$ &$0.22$ &$0.10$ &$0.10$\\\hline
$\alpha_s^{\rm PQCD}(q^2)$&---& $0.584$&$0.354$ &  $0.218$ &  $0.118$ \\
$\alpha_s^{\rm C}(q^2)$&$0.644^{+0.877}_{-0.177}$&$0.532^{+0.147}_{-0.096}$&$0.357^{+0.022}_{-0.027}$&$0.212^{+0.001}_{-0.002}$ & $0.120^{+0.000}_{-0.000}$ \\
\hline\hline
\end{tabular}
\end{center}
\end{table}

\tb{In Table~\ref{tab:fit}, we summarize the results of $\alpha_s$ at different matching point of $q^2$. In addition, in order to further test the values of $\Lambda_{\rm C}$ given above, the values of $\alpha_s$ at large $q^2=m_Z^2$ and freezing point $q^2=0$ are also listed in the Table~\ref{tab:fit}.  It can be found that: (i) Using the $\Lambda_{\rm C}=0.1~{\rm GeV}$ fitted at $q^2=m_b^2$, our prediction $\alpha_s^{\rm C}(m_Z^2)=0.120$ is in agreement with the $\alpha_s^{\rm PQCD}(m_Z^2)$ and the experimental data $0.1182$~\cite{PDG2016}. Moreover, the uncertainty induced by the $m_{g}=0.5\pm0.2~{\rm GeV}$ vanishes at  $q^2=m_Z^2$, but is very large at small $q^2$ region. (ii) Our prediction for the freezing value, $\alpha_s^{\rm C}(0)=0.644^{+0.877}_{-0.177}$,  is also in agreement with, for instance, $\alpha_s^{\rm LFH}(0)=1.22\pm0.04\pm0.11\pm0.09 $~\cite{Deur:2016cxb}  obtained in the framework of the light-front holographic QCD and  $\overline{\rm MS}$ RS, within the theoretical uncertainties.
}

Using above formulae and the same convention as scheme~I, we obtain the annihilation amplitudes,
{\setlength\arraycolsep{1pt}
\begin{eqnarray}
A_{1}^{i,0}(V_{1},V_{2}) = \pi\int_0^1\! dx dy  \,\alpha_{s}^{\rm C}(q^{2})\, \Bigg\{\Phi_{V_{1}}(x)\,\Phi_{V_{2}}(y)\,\bigg[\frac{\bar{y}}
{(x\bar{y}-\omega^{2}(q^2)+i\epsilon)\big[(x+\bar{y})z_{b}-{x}\bar{y}-i\epsilon\big]} \nonumber\\[0.1cm]
- \frac{x}{(x\bar{y}-\omega^{2}(q^2)+i\epsilon)\big[({x}+\bar{y})z_{c}-{x}\bar{y}-i\epsilon\big]}\bigg] \nonumber \\[0.2cm]
 - r_{\chi}^{V_{1}}\,r_{\chi}^{V_{2}}\,\Phi_{v_{1}}(x)\,\Phi_{v_{2}}(y)\, \bigg[\frac{{x}\bar{y}+(x+\bar{y}-2x\bar{y})z_{b}}
 {(x\bar{y}-\omega^{2}(q^2)+i\epsilon)\big[({x}+\bar{y})z_{b}-{x}\bar{y}-i\epsilon\big]}
 \nonumber\\[0.1cm]
 - \frac{{x}\bar{y}+(x+\bar{y}-2x\bar{y})z_{c}}
 {(x\bar{y}-\omega^{2}(q^2)+i\epsilon)\big[({x}+\bar{y})z_{c}-{x}\bar{y}-i\epsilon\big]}\bigg]\Bigg\}\,,\nonumber\\[0.1cm]
 \label{eq:a1i0_schemeII}
\end{eqnarray}
\begin{eqnarray}
A_{1}^{i,-}(&V_{1}&,V_{2})
=\pi\frac{2m_{1}m_{2}}{m_{B_{c}}^2}\int_0^1\! dx dy  \,\alpha_{s}^{\rm C}(q^{2})\,
\phi_{b1}(x)\,\phi_{b2}(y)\,
\nonumber \\[0.2cm]
& &\Bigg[\bigg(\frac{\bar{y}+z_{b}}
{(x\bar{y}-\omega^{2}(q^2)+i\epsilon)\big[(x+\bar{y})z_{b}-{x}\bar{y}-i\epsilon\big]}
+ \frac{\bar{x}-z_{c}}{(x\bar{y}-\omega^{2}(q^2)+i\epsilon)\big[({x}+\bar{y})z_{c}-{x}\bar{y}-i\epsilon\big]}\bigg)
 \nonumber \\[0.2cm]
 &+&\bigg(\frac{\bar{x}\bar{y}^2}{(x\bar{y}-\omega^{2}(q^2)+i\epsilon)^2\big[({x}+\bar{y})z_{b}-{x}\bar{y}-i\epsilon\big]}
 - \frac{\bar{x}\bar{y}^2}{(x\bar{y}-\omega^{2}(q^2)+i\epsilon)\big[(x+\bar{y})z_{b}-x\bar{y}-i\epsilon\big]^2}\bigg)
\nonumber \\[0.2cm]
&-& \bigg(\frac{x\bar{x}\bar{y}}{(x\bar{y}-\omega^{2}(q^2)+i\epsilon)^2\big[({x}+\bar{y})z_{c}-{x}\bar{y}-i\epsilon\big]}
 -\frac{x\bar{x}\bar{y}}{(x\bar{y}-\omega^{2}(q^2)+i\epsilon)\big[(x+\bar{y}){z_{c}}-x\bar{y}-i\epsilon\big]^2}\bigg)\Bigg]\,,\nonumber\\[0.1cm]
 \label{eq:a1im_schemeII}
\end{eqnarray}
\begin{eqnarray}
A_{1}^{i,+}(&V_{1}&,V_{2}) = \pi\frac{2m_{1}m_{2}}{m_{B_{c}}^2}\int_0^1\! dx dy  \,\alpha_{s}^{\rm C}(q^{2})\,
\phi_{a1}(x)\,\phi_{a2}(y)\,
\nonumber \\[0.2cm]
& &\Bigg[\bigg(\frac{z_{b}-y}
{(x\bar{y}-\omega^{2}(q^2)+i\epsilon)\big[(x+\bar{y})z_{b}-{x}\bar{y}-i\epsilon\big]}
- \frac{x+z_{c}}{(x\bar{y}-\omega^{2}(q^2)+i\epsilon)\big[(x+\bar{y})z_{c}-{x}\bar{y}-i\epsilon\big]}\bigg)
 \nonumber \\[0.2cm]
 &+&\bigg(\frac{xy\bar{y}}{(x\bar{y}-\omega^{2}(q^2)+i\epsilon)^2\big[({x}+\bar{y})z_{b}-{x}\bar{y}-i\epsilon\big]}
 - \frac{xy\bar{y}}{(x\bar{y}-\omega^{2}(q^2)+i\epsilon)\big[(x+\bar{y})z_{b}-x\bar{y}-i\epsilon\big]^2}\bigg)
\nonumber \\[0.2cm]
&-& \bigg(\frac{x^2y}{(x\bar{y}-\omega^{2}(q^2)+i\epsilon)^2\big[(x+\bar{y})z_{c}-{x}\bar{y}-i\epsilon\big]}
 -\frac{x^2y}{(x\bar{y}-\omega^{2}(q^2)+i\epsilon)\big[(x+\bar{y}){z_{c}}-x\bar{y}-i\epsilon\big]^2}\bigg)\Bigg]\,,\nonumber\\[0.1cm]
 \label{eq:a1ip_schemeII}
\end{eqnarray}
where, $\omega^{2}(q^2)=M_{g}^{2}(q^{2})/m_{B_c}^2$ with $q^2 \simeq x\bar{y}m_{B_{c}}^2$ the time-like gluon momentum square.
\tb{Here, we would like to clarify that the IR finite gluon propagator given by Eq.~(\ref{eq:gluonprop}) is used for evaluating both twist-3 and twist-2 contributions for consistence, although it is not essential for the latter from the viewpoint of regulating end-point divergence~(the integral of twist-2 part is finite as has been mentioned in the last subsection).}
Again we checked that the results for $B_{u,d,s }\to  V V$ decay with  IR finite gluon propagator, which have been calculated in  Ref.~\cite{Chang:2012xv}, can be recovered from above formulae by taking the limits $z_b\to 1$ and $z_c\to 0$.

From Eqs.~(\ref{eq:a1i0_schemeII}), (\ref{eq:a1im_schemeII}) and (\ref{eq:a1ip_schemeII}), it is found that the singularities of the gluon propagators are moved from end-point into integral intervals by using the infrared finite form of the gluon propagator. \tb{Using $m_{g}=0.5~{\rm GeV}$}, we obtain the numerical results for the building blocks $A_{1}^{i,\lambda}(V,V)$ in scheme II that
\tb{
\begin{eqnarray}
 A_{1}^{i,0}(V_{1},V_{2}) & = & \pi \left[(-10.33 - 5.21i) - r_{\chi}^{V_{1}}r_{\chi}^{V_{2}}(-1.40 - 8.75i)\right],  \\[0.2cm]
 A_{1}^{i,-}(V_{1},V_{2}) & = &\pi \frac{2\,m_{1}\,m_{2}}{m_{B_{c}}^2}(16.82 - 2.13i),\\[0.2cm]
 A_{1}^{i,+}(V_{1},V_{2}) & = &\pi \frac{2\,m_{1}\,m_{2}}{m_{B_{c}}^2}(-17.33 + 8.18i).
 \label{eq:a1i_schemeII_num1}
\end{eqnarray}
}
Comparing with the results in scheme I, we find that the annihilation contributions are enhanced when we adopt the infrared finite gluon propagator.

\tb{It is known that the form of IR finite gluon propagator, Eqs.~(\ref{eq:cornwall1}) and (\ref{eq:effmass}), is model-dependent. In order to  estimate the model-dependence of scheme II, we would like to reevaluate the annihilation amplitude by using  Aguilar-Papavassiliou~(AP)'s prescription~\cite{Aguilar:2007ie} instead of Cornwall's solution. The relevant formulae and inputs are collected in Appendix~\ref{appendix C}.  For simplicity, we take $A_{1}^{i,0}(\rho^{-}\omega)$ as an example. The result evaluated by using Cornwall's solution with $m_g=0.5\pm0.2 {\rm GeV}$ are also shown in the follows for convenience of comparison. Numerically, we obtain
\begin{eqnarray}
|A_{1}^{i,0}(\rho^{-}\omega)|=32.56~({\rm AP})\quad {\rm vs.} \quad 36.36_{-4.47}^{+6.73}~({\rm Cornwall}).
\end{eqnarray}
It can be clearly seen that these results are consistent with each other within the uncertainties of $m_g$ ({\it i.e.}, the possible model-dependence of scheme II could be accommodate by using a conservative input, $m_g=0.5\pm0.2 {\rm GeV}$ ).}


\section{Numerical results and discussions}
\label{sec3}
Using the building blocks given in the last section, we summarize the polarization-dependent decay amplitudes ${\cal A}_{\lambda}(B_{c}^{-}\to V_1V_2)$ in the Appendix~\ref{appendix A}.
The branching ratios for charmless $B_{c}^{-} \to V_{1}V_{2}$ decays in the rest frame of $B_{c}^{-}$ meson can be written as
\begin{equation}
 \tb{{\cal B}}(B_{c}^{-}\to V_1V_2) = \frac{\tau_{B_c}}{8\pi}\,\frac{|\vec{p}|}{m_{B_c}^2}\sum_{\lambda} \big|{\cal A}_{\lambda}(B_{c}^{-}\to V_1V_2)\big|^2\,,
\end{equation}
where \tb{$\tau_{B_c}=0.507~\mathrm{ps}$}\,\cite{PDG2016} is the  lifetime of $B_c$-meson, and $|\vec{p}|$ is the center-of-mass momentum of either of the two outgoing mesons,
\begin{equation}
|\vec{p}|=\frac{\sqrt{\left[m_{B_c}^2-(m_{V_1}+m_{V_2})^2\right]\left[m_{B_c}^2-(m_{V_1}-m_{V_2})^2\right]}}{2 m_{B_c}}\,.
\end{equation}
Besides of the branching fraction, the polarization fractions defined as
\begin{equation}
f_{L,\parallel,\perp}=\frac{\big|{\cal A}_{0,\parallel,\perp}\big|^2}
{\big|{\cal{A}}_{0}\big|^2+\big|{\cal{A}}_{\parallel}\big|^2+\big|{\cal{A}}_{\perp}\big|^2}
\end{equation}
are also very important observable, where ${\cal A}_{\parallel}$ and ${\cal A}_{\perp}$ are parallel and perpendicular amplitudes and could be easily gotten through ${\cal A}_{\parallel,\perp}=({\cal A}_{-}\pm{\cal A}_{+})/\sqrt{2}$. In addition,  the CP-violating asymmetries for all of the decay modes considered in this paper are absent.

\begin{table}[t]
 \centering
 \tabcolsep 0.12in
 \tb{\caption{\label{tab02} \small The CP-averaged branching ratios~(in units of $10^{-8}$ ) of $B_{c}\to VV$ decays based on the two regulation schemes. The theoretical errors correspond to the uncertainties induced by ``CKM", ``hadronic", ``scale", and ``$m_g$". The pQCD predictions~\cite{Liu:2009qa} are also listed in the last column.}}
 \begin{center}
 \begin{tabular}{|l|c|l|l|l|}
 \hline\hline
 \multicolumn{1}{|c|}{Decay modes} & Cases & Scheme~I & Scheme~II & pQCD \\
 \hline\hline
 $B_{c}^{-}\to\rho^{-}\rho^{0}$ & $|\Delta S|=0 $
 & $0$
 & $0$
 & $0$\\

 $B_{c}^{-}\to\rho^{-}\omega$ & $|\Delta S|=0 $
 & ${16.1}^{\,+0.8\,+ 2.9\,+50.3}
 _{\,- 1.1\,- 2.5\,-12.0}$
 & ${30.6}^{\,+ 1.6\,+ 5.6\,+41.7\,+ 14.3}
 _{\,- 2.1\,- 4.7\,-18.7\,- 6.1}$
 & $106^{+32+21}_{-2-2}$\\

 $B_{c}^{-}\to K^{*-}K^{*0}$& $|\Delta S|=0$
 & ${9.21}^{\,+0.47\,+1.39\,+28.79}_{\,-0.63\,-1.16\,- 6.86}$
 & ${17.5}^{\,+ 0.9\,+ 2.7\,+23.9\,+ 8.8}
 _{\,- 1.2\,- 2.2\,-10.7\,- 3.5}$
 & $100^{+6+81}_{-4-48}$\\[0.2cm]

 $B_{c}^{-}\to  K^{*-}\rho^{0}$& $|\Delta S|=1 $
 & ${0.26}^{\,+0.01\,+0.04\,+0.82}_{\,-0.02\,-0.03\,-0.19}$
 & ${0.50}^{\,+0.03\,+0.07\,+0.68\,+0.24}_{\,-0.04\,-0.06\,-0.31\,-0.10}$
 & $3^{+0+1}_{-0-1}$\\

 $B_{c}^{-}\to \bar{K}^{0*}\rho^{-}$& $|\Delta S|=1 $
 & ${0.53}^{\,+0.02\,+0.07\,+1.64}_{\,-0.04\,-0.07\,- 0.40}$
 & ${1.00}^{\,+ 0.05\,+ 0.14\,+1.36\,+ 0.48}
 _{\,- 0.07\,- 0.12\,- 0.61\,- 0.20}$
 & $6^{+0+2}_{-0-1}$\\

 $B_{c}^{-}\to K^{*-}\omega$& $|\Delta S|=1 $
 & ${0.20}^{\,+0.02\,+0.02\,+0.64}_{\,-0.01\,-0.03\,-0.15}$
 & ${0.39}^{\,+0.02\,+0.07\,+0.53\,+0.18}_{\,-0.03\,-0.07\,-0.24\,-0.08}$
 & $3^{+0+0}_{-0-2}$\\

 $B_{c}^{-}\to \phi K^{*-}$& $|\Delta S|=1 $
 & ${0.58}^{\,+0.04\,+0.06\,+1.83}_{\,-0.04\,-0.05\,- 0.43}$
 & ${1.11}^{\,+ 0.05\,+ 0.12\,+1.51\,+ 0.57}
 _{\,- 0.08\,- 0.10\,- 0.68\,- 0.22}$
 & $5^{+0+1}_{-1-3}$\\
 \hline\hline
\end{tabular}
\end{center}
\end{table}

With the theoretical formulae given above and the input parameters collected in Appendix~\ref{appendix B}, we proceed to present the numerical results for  the CP-averaged branching ratios of $B_c\to VV$ decays.
In our calculation, the default value of renormalization scale is set at $\mu=m_{B_c}/2$, which is approximately the averaged virtuality of the time-like gluon propagated in the annihilation diagrams. The numerical results based on the two schemes for regulating the end-point divergence are collected in Table~\ref{tab02}. In this table, we present the ``default results" along with the detailed errors estimated with different theoretical uncertainties of inputs. The first error refers to the variation of the CKM parameters (named as ``CKM"); the second error corresponds to the quark masses and decay constants~(named as ``hadronic"); the third error originates from the variation of the renormalization scale $\mu$~(named as ``scale"); the last error  in scheme II reflects the uncertainty of the effective gluon mass $m_g$~(named as ``$m_g$"). For comparison, \tb{ predictions of the pQCD factorization approach~\cite{Liu:2009qa} are also listed in the last column of Table~\ref{tab02}~(\tb{the $\overline{\rm MS}$ RS is used in the pQCD calculation, see Refs.~\cite{Liu:2009qa,Lu:2000em} for detail}).}

Based on the results collected in Table~\ref{tab02}, we have the following observations and remarks:
\begin{itemize}
\item From the decay amplitudes summarized  in the Appendix~\ref{appendix A}, it can be found that the CKM matrix elements and Wilson coefficients are the key factors to determine the size of the amplitude. The strangeness-changing~($|\triangle S|=1$) processes are CKM-suppressed relative to the strangeness-conserving~($|\triangle S|=0$) processes due to the hierarchy of the CKM matrix elements, $|V_{ud}/V_{us}|^2\sim 19$. As a result, the branching ratios of $|\triangle S|=0$ decay channels are generally much larger than those of $|\triangle S|=1$ ones by about an order of magnitude.

\item Because the contributions from $u\bar{u}$ and $d\bar{d}$ components of the $\rho^{0}$ meson, $|\rho^0\rangle=(|u\bar{u}\rangle-|d\bar{d}\rangle)/\sqrt{2}$, cancel with each other exactly, the branching ratio of $B_{c}^{-}\to\rho^{-}\rho^{0}$ decay is \tb{zero}. On the other hand, for $B_{c}^{-}\to\rho^{-}\omega$ decay, the interference between the two flavor components $u\bar{u}$ and $d\bar{d}$ of the $\omega$ meson is constructive due to $|\omega\rangle=(|u\bar{u}\rangle+|d\bar{d}\rangle)/\sqrt{2}$, which results in a large branching ratio.

 \item Among the charmless $B_{c} \to VV$ decays considered in this work, the  CKM-favored $B_{c}^{-}\to\rho^{-}\omega$, $K^{*-}K^{*0}$ decay modes have relatively large branching ratios being around ${\cal O}(10^{-7})$.  It is also found that $B_{c}^{-}\to \rho^{-}\omega$ decay has the the largest branching ratio,   \tb{$\sim 30.6\times10^{-8}$}, and hence will possibly be observed earlier at LHC and SuperKEKB.

\item
In the limit of the $ SU(3)$ flavor-symmetry, the relation
  \begin{equation}\label{eq:su3}
     {\cal A}(B_{c}^{-}\to \bar{K}^{0*}\rho^{-})=\sqrt{2}{\cal A}(B_{c}^{-}\to K^{*-}\rho^{0})=\hat{\lambda}{\cal A}(B_{c}^{-}\to K^{*-}K^{*0})\,,
     \end{equation}
 with the Cabibbo-suppressing factor $\hat{\lambda}=V_{us}/V_{ud}$ is expecetd.  In our calculation, we take the asymptotic expressions for the distribution amplitudes of $V$ mesons, and the flavor-asymmetry effect arises only from the chiral enhancement parameter $r_{\chi}^{V}$ and decay constants. Therefore, the $SU(3)$ breaking effect turns out to be relatively small, and above relation still holds approximately in \tb{both scheme I and II} which can be seen  from Table~\ref{tab02}.


\item \tb{ In scheme II, using the central values of input parameters summarized  in Appendix~\ref{appendix B} and the AP's prescription~\cite{Aguilar:2007ie}  given in Appendix~\ref{appendix C} instead of Eqs.~(\ref{eq:cornwall1}) and (\ref{eq:effmass}), we have calculated the branching ratios of the $|\triangle S|=0$ processes, and obtain
 \begin{equation}
{\cal B}(B_{c}^{-}\to\rho^{-}\omega)=25.0\times10^{-8},\qquad
{\cal B}(B_{c}^{-}\to K^{*-}K^{*0})=14.5\times10^{-8}\,.
 \end{equation}
Comparing with corresponding results in Table~\ref{tab02}, one can find that such results are in agreement with the ones obtained  by using Cornwall's formulae, Eqs.~(\ref{eq:cornwall1}) and (\ref{eq:effmass}), within the theoretical uncertainties of $m_g=0.5\pm0.2 {\rm GeV}$.
}

\begin{figure}[t]
\includegraphics[width=8.0cm]{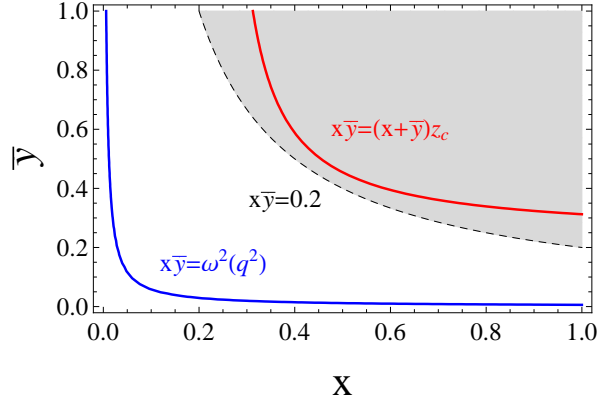}
\caption{\tb{The singularities in the integrals induced by the c-quark mass~(red line) and the dynamical gluon mass~(blue line). The dashed line and shaded region correspond to $ x\bar{y}=0.2$ and  $0.2\leqslant x\bar{y}\leqslant  1$, respectively (see text for explanation).}}
\label{fig.S}
\end{figure}
\item
\tb{As mentioned in the last section, the annihilation contributions are enhanced when we adopt the IR finite gluon propagator. It is mainly caused by that:
(i) In scheme I, the strong coupling in the amplitude is determined by the scale $\mu$ with the default value $m_{B_c}/2$; while,  $\alpha_{s}(q^2)$ in scheme II is determined by Eq.~(\ref{eq:cornwall1}), and relatively larger than the one in scheme I at low $q^2$ region.
(ii) It have been found that the singularities in the integral interval can significantly affect the numerical results of the integrals in the annihilation amplitudes, for instance, Ref~\cite{Chang:2008tf}. In scheme I, the singularities $x\bar{y}=(x+\bar{y})z_c$ is induced by the sizable $c-$quark mass; while,  in scheme II, besides $x\bar{y}=(x+\bar{y})z_c$, additional singularities $x\bar{y}=\omega^2(q^2)$ induced by effective gluon mass enter into the integral interval~( the singularities are shown in Fig.~\ref{fig.S} ). As a result, the branching ratios in scheme II are generally larger than the ones in scheme I, but they are still in agreement within the large theoretical uncertainties.
}

\tb{In fact, schemes I and II result in similar annihilation contributions at large $q^2$ region. In order to clearly show that, we take the integral interval $0.2\leqslant x\bar{y}\leqslant  1$, which is far from the the singularities $x\bar{y}=\omega^2(q^2)$ as Fig.~\ref{fig.S} shows; and take the amplitude of  twist-2 part in $A_1^{i,0}$ as an example,  which  plays a dominative role  in the annihilation amplitudes. Then, using the central values of the other inputs, we obtain
 \begin{equation}
|A_1^{i,0}(\text{twist-2})|_{0.2\leqslant x\bar{y}\leqslant  1}=20.1~({\rm scheme~I})\quad {\rm vs.} \quad 22.9~({\rm scheme~II})\,,
 \end{equation}
 which are similar to each other, and therefore, confirm our analyses given above.}

\item \tb{From Table~\ref{tab02}, we find that our predictions~(central value) are relatively smaller than the ones in the pQCD factorization approach~\cite{Liu:2009qa}.
The different choices of the renormalization scale and  strategies for coping with the end-point contributions may be the main reasons leading to these discrepancies.}

\tb{
For the $B_{c}^{-}\to \bar{K}^{0*}\rho^{-}$ and $ K^{*-}\rho^{0}$ decays induced by the strangeness-changing~($|\triangle S|=1$) transition, the relation
\begin{equation}\label{eq:rela}
{\cal B}(B_{c}^{-}\to \bar{K}^{0*}\rho^{-})= 2\times {\cal B} (B_{c}^{-}\to  K^{*-}\rho^{0})
 \end{equation}
 is expected by  $ SU(3)$ flavor-symmetry as Eq.~(\ref{eq:su3}) shows, and also can be clearly seen from Eqs.~(\ref{eq:kmrho}) and (\ref{eq:kzrho}).
It can be found from Table~\ref{tab02} that such relation is favored by the predictions of this work and pQCD approach.
 For the strangeness-conserving~($|\triangle S|=0$) processes, the amplitudes, Eqs. (\ref{eq:rhomw}) and (\ref{eq:kk}), imply a rough relation similar to Eq.~(\ref{eq:rela}), ${\cal B}(B_{c}^{-}\to\rho^{-}\omega)\,\sim\,2\times\,{\cal B}(B_{c}^{-}\to K^{*-}K^{*0})$. It is satisfied by our results but disfavored by pQCD, which can be seen from  Table~\ref{tab02}.    }


\tb{
The heavy flavor experiments at LHC and SuperKEKB/Belle-II in the future are expected to exhibit a clear picture for the annihilation contributions.
 }

\item  As we have mentioned, because only the twist-3 terms of the light-cone projector for the final-state $V$ mesons contribute to the transverse amplitudes, one can find  from Eqs.~(\ref{eq:a1i0_schemeI}-\ref{eq:a1ip_schemeI}) and (\ref{eq:a1i0_schemeII}-\ref{eq:a1ip_schemeII}) that $A_{1}^{i,\pm}$ are suppressed by two powers of $\Lambda_{QCD}/m_{B_{c}}$ compared with $A_{1}^{i,0}$. As a result, all of the $B_{c}\to VV$ decays  are dominated by longitudinal polarization. Numerical, we obtain $f_L(B_{c}\to VV)\gtrsim 99\%$, which is a little larger than the pQCD prediction $f_L(B_{c}\to VV)\sim[86,95]\%$~\cite{Liu:2009qa} and will be tested by the future measurements.


  \end{itemize}


\section{Summary}
\label{sec4}
In this paper, we have studied the nonleptonic charmless $B_{c}\to VV$ decays within the framework of QCD factorization.
These decay modes can occur only via the weak annihilation diagram, which involves only a tree operator, $Q_2$, at the order of $\alpha_s$, and therefore,  they will provide an important testing ground for the magnitude of annihilation contribution and the underlying mechanism.

It is found that the transverse amplitudes and the twist-3 part of longitudinal amplitude are power-suppressed by $(\Lambda_{QCD}/m_{B_{c}})^2$ relative to the main contribution~(the twist-2 part of longitudinal amplitude),  and are \tb{small} numerically. For the main contribution, the problem of end-point divergence, which appeared in the  $B_{u,d,s}\to MM$ decays, vanishes in the $B_{c}\to VV$ decays due to the sizable $c$-quark mass. However, it still exists in the power-suppressed corrections. In order to regulate the end-point divergence, we have employed two different schemes based on the parameterization and the infrared finite gluon propagator with a dynamical gluon mass, respectively. Our predictions for the branching fractions of $B_{c}\to VV$ decays are collected in Table~\ref{tab02}, in which  the $B_{c}^{-}\to\rho^{-}\omega$ and $K^{*-}K^{*0}$ decays have relatively large branching ratio, $\sim 10^{-7}$, and hence have the best potential for the detection. In addition, the longitudinal polarization fractions are expected at the level of $99\%$ for all of the $B_{c}\to VV$ decays. Then, some phenomenological analyses and discussions are made.  All of the findings in this paper are waiting for the experimental test at LHC and SuperKEKB/Belle-II in the future.

\begin{appendix}

\section{Decay amplitudes of $B_c\to VV$ decays}
\label{appendix A}

Starting with Eq.~(\ref{eq:matrixelement}) and adopting the standard phase convention for the flavor wavefunctions of light and heavy mesons~\cite{Beneke:2001ev,Beneke:2002jn,Beneke:2006hg}, one can easily write down the decay amplitude for a given decay mode.
There are seven charmless $B_c\to VV$ decays with the corresponding amplitude given, respectively, as~(the exact isospin symmetry is assumed):
\begin{eqnarray}
{\cal A}^{\lambda}(B_{c}^{-}\to\rho^{-}\rho^{0}) &=& \frac{G_{F}}{2}\,V_{cb}V_{ud}^{*}\,f_{B_{c}}f_{\rho^{-}}
f_{\rho^{0}} \left[b_{2}^{\lambda}(\rho^{0}, \rho^{-})-b_{2}^{\lambda}(\rho^{-}, \rho^{0})\right]=0\,,\\[0.2cm]
\label{eq:rhomw}
{\cal A}^{\lambda}(B_{c}^{-}\to\rho^{-}\omega) &=& \frac{G_{F}}{2}\,V_{cb}V_{ud}^{*}\,f_{B_{c}}f_{\rho^{-}} f_{\omega} \left[b_{2}^{\lambda}(\omega, \rho^{-})+b_{2}^{\lambda}(\rho^{-}, \omega)\right]\,,\\[0.2cm]
\label{eq:kk}
{\cal A}^{\lambda}(B_{c}^{-}\to K^{*-}K^{*0}) &= &\frac{G_{F}}{\sqrt{2}}\,V_{cb}V_{ud}^{*}\,f_{B_{c}}f_{K^{*-}}f_{K^{*0}}\, b_{2}^{\lambda}(K^{*-},K^{*0})\,, \\[0.2cm]
\label{eq:kmrho}
{\cal A}^{\lambda}(B_{c}^{-}\to  K^{*-}\rho^{0}) &=& \frac{G_{F}}{2}\,V_{cb}V_{us}^{*}\,f_{B_{c}}f_{K^{*-}}f_{\rho^{0}}\, b_{2}^{\lambda}(\rho^{0},K^{*-})\,, \\[0.2cm]
\label{eq:kzrho}
{\cal A}^{\lambda}(B_{c}^{-}\to \bar{K}^{0*}\rho^{-}) &=& \frac{G_{F}}{\sqrt{2}}\,V_{cb}V_{us}^{*}\,f_{B_{c}} f_{\bar{K}^{0*}}f_{\rho^{-}}\,b_{2}^{\lambda}(\rho^{-},\bar{K}^{0*})\,, \\[0.2cm]
{\cal A}^{\lambda}(B_{c}^{-}\to K^{*-}\omega) &=& \frac{G_{F}}{\tb{2}}\,V_{cb}V_{us}^{*}\,f_{B_{c}}
f_{K^{*-}}f_{\omega}\,b_{2}^{\lambda}(\omega,K^{*-})\,, \\[0.2cm]
{\cal A}^{\lambda}(B_{c}^{-}\to \phi K^{*-}) &=& \frac{G_{F}}{\sqrt{2}}\,V_{cb}V_{us}^{*}\,f_{B_{c}}
f_{\phi}f_{ K^{*-}}\,b_{2}^{\lambda}( K^{*-},\phi)\,.
\end{eqnarray}

\section{Input parameters}
\label{appendix B}

For the CKM matrix elements, we adopt the Wolfenstein parameterization~\cite{Wolfenstein:1983yz} and keep terms up to $\mathcal{O}(\lambda^4)$ :
\begin{eqnarray}
V_{ud}=1-\frac{1}{2}\lambda^2-\frac{1}{8}\lambda^4+\mathcal{O}(\lambda^6)\,, \qquad
V_{us}=\lambda+\mathcal{O}(\lambda^7)\,, \qquad V_{cb}=A\lambda^2+\mathcal{O}(\lambda^8)\,,
\end{eqnarray}
with the inputs $A=0.811\pm0.026$ and $\lambda=0.22506\pm0.00050$~\cite{PDG2016}.

The hadronic inputs are summarized as follows: the pole masses of quarks are~\cite{PDG2016}
\begin{eqnarray}
 m_u=m_d=m_s=0, \quad m_c = 1.67 \pm 0.07~\mathrm{GeV}, \nonumber\\
 m_b = 4.78 \pm 0.06~\mathrm{GeV}\,,\quad  m_t=174.2\pm1.4~{\rm GeV}\,;
\end{eqnarray}
In the evaluation of Wilson coefficients $C_i(\mu)$~\cite{Buras:1998raa}, the following inputs~\cite{PDG2016},
\begin{eqnarray}
&& \alpha_s(M_Z)=0.1182\pm0.0012,
\qquad \sin^2\theta_W=0.2313, \nonumber \\
&& M_Z=91.1876~{\rm GeV}, \quad M_W=80.385~{\rm GeV}, \quad
\end{eqnarray}
are used. For the decay constants of light mesons, we take the results given in Ref.~\cite{Jung:2012vu}, which are an update of the ones extracted in Ref.~\cite{Ball:2006eu}; for the decay constant of $B_c$ meson, we adopt the results based on the lattice QCD~\cite{Chiu:2007km}. Their values are
\begin{eqnarray}
&& f_{\rho} = 215\pm 6~\mathrm{MeV},\qquad f_{\rho}^\perp(2~{\rm GeV})/f_{\rho} = 0.70\pm0.04,\nonumber\\
&& f_{K^{*}} = 209\pm 7~\mathrm{MeV},\quad  f_{K^{*}}^\perp(2~{\rm GeV})/f_{K^{*}} = 0.73\pm0.04,\nonumber\\
&& f_{\omega} = 188\pm 10~\mathrm{MeV},\quad
f_{\omega}^\perp(2~{\rm GeV})/f_{\omega} = 0.70\pm0.10,\nonumber\\
&& f_{\phi} = 229\pm 3~\mathrm{MeV},\qquad
  f_{\phi}^\perp(2~{\rm GeV})/f_{\phi} = 0.750\pm0.020\quad\nonumber\\
&& f_{B_c} = 487 \pm 5~\mathrm{MeV},\quad
\end{eqnarray}
in which, the scale dependence of the transverse decay constants is taken into account via the leading-logarithmic running $f_\perp(\mu) = f_\perp(\mu_0) \,\left[\alpha_s(\mu)/\alpha_s(\mu_0)\right]^{4/23}$~\cite{Ball:2006eu}. \tb{For the renormalization scale, we take $\mu=m_{B_c}/2$ as default input, and vary it in the range $[m_{B_c}/2,m_{B_c}]$
to assess the scale  uncertainty. }

\tb{For the other well-determined inputs, such as the masses and lifetimes of mesons and Fermi constant {\it etc.}, we take their central values given by PDG~\cite{PDG2016}. In addition, the values of specific parameters, $\Lambda_h$ in scheme I and $m_g$ in scheme II, are given and discussed in the text.  }


\section{{Aguilar-Paparassiliou's prescription}}
\label{appendix C}

\tb{ Besides Cornwall's prescription,   Aguilar and Paparassiliou also show a massive propagator effectively describing the solution of the Schwinger-Dyson equation with two possible behaviors at large $q^2$~\cite{Aguilar:2007ie}. One is similar to Eq. (\ref{eq:effmass}), with logarithmic running $m_g (q^2 )\propto \ln(q^2 )^{-\gamma}$~\cite{Aguilar:2007ie}. The second power-law solution has the form
 \begin{equation}
 m^2(q^2)=\frac{m^4_0}{q^2+m^2_0}\Bigg[\ln
\left(\frac{q^2+\rho\,m^2_0}{\Lambda^2}\right)\Big/\ln\left(\frac{\rho\,m^2_0}{\Lambda^2}\right) \Bigg]^{\gamma_2-1} \,,
\end{equation}
which is used in the evaluation to compare with the results based on the Cornwall's prescription.
The running coupling constant is given by
\begin{equation}
g^2(q^2) = \bigg[ {b}\ln\left(\frac{q^2 + f(q^2, m^2(q^2))}{\Lambda^2}\right)\bigg]^{-1}\,,
\end{equation}
where the function $f(q^2, m^2(q^2))$ is given by a power law expression
\begin{equation}
f(q^2, m^2(q^2)) = \rho_{\,1} m^2(q^2)+ \rho_{\,2} \frac{m^4(q^2)}{q^2+m^2(q^2)} +\rho_{\,3} \frac{m^6(q^2)}{[q^2+m^2(q^2)]^{\,2}} \,,
\end{equation}
with $b=33/48\pi^2$. The values of parameters $\rho=1.046$, $m_{0}^2=0.5$Ge$V^2$, $\Lambda=0.3 \mathrm{GeV}$, $\gamma_2=2.12$, $\rho_1=1.205$, $\rho_2=-0.690$, $\rho_{3}=0.121$ are suggested in Ref.~\cite{Aguilar:2007ie}. One may refer to Refs.~\cite{Aguilar:2007ie,Deur:2016tte} for detail for this part. }

\end{appendix}


\section*{Acknowledgements}
This work is supported by the National Natural Science Foundation of China (Grant Nos. 11475055, 11547014 and 11275057). Q.~Chang is also supported by the Foundation for the Author of National Excellent Doctoral Dissertation of China (Grant No. 201317), the Program for Science and Technology Innovation Talents in Universities of Henan Province (Grant No. 14HASTIT036), the Excellent Youth Foundation of HNNU and the CSC (Grant No. 201508410213).


\end{document}